\documentclass{ws-ijmpcs}

\begin{document}

\markboth{N.K. Walford and F.J. Klein}
{Polarization Observables for $K^{+}\Lambda$ and $K^{+}\Sigma^{0}$ Photoproduction from Polarized Protons}

%
\catchline{}{}{}{}{}
%

\title{Polarization Observables for $K^+\Lambda$ and $K^{+}\Sigma^{0}$ Photoproduction from Polarized Protons}

\author{Natalie K. Walford and Franz J. Klein (for the CLAS Collaboration)}

\address{Physics Department, The Catholic University of America\\
Washington, DC 20064,
USA\\
natalie@jlab.org}

\maketitle

\begin{history}
\received{03 August 2013}
\end{history}

\begin{abstract}
The search for undiscovered excited states of the nucleon continues to be a focus of experiments at Jefferson Lab. Recent LQCD calculations have confirmed long-standing quark-model predictions of many more states than have so far been identified.\cite{lqcd} A large effort for the N$^\ast$ program has been launched using the CLAS detector to provide the database that will allow nearly model-independent partial wave analyses to be carried out in the search for such states. Polarization observables play a crucial role in this effort, as they are essential in disentangling overlapping resonant and non-resonant amplitudes. 
In 2010, double-polarization data were taken at JLab using circularly polarized photons incident on a transversely polarized frozen-spin butanol target.\cite{frost} Our current analysis yields preliminary data of the $T$ and $F$ asymmetries of the $K^{+}\Lambda$ and $K^{+}\Sigma^{0}$ final states, which are compared to predictions of recent multipole analyses. 
This work is the first of its kind and will significantly broaden the world database for these reactions.

\end{abstract}

\ccode{PACS numbers: 13.60.Le, 25.20.Lj, 13.88.+e}

\section{Introduction}

The nucleon spectrum can provide a tool in understanding the dynamics and relevant degrees of freedom within hadrons in the non-perturbative regime of QCD. QCD describes the strong force acting between quarks, but only leads to solutions at very low energies and high energies where perturbative methods can be used.\cite{proposal} At medium energies, nonperturbative approximation models are required to describe the behavior of quarks in nucleons. In particular, constituent quark models of baryons are used to determine the spectrum of resonant states of the nucleon. These models are based on the symmetric behavior of the three valence quarks and predict far more resonant states than have been observed. The non-observation of predicted resonances may be explained by assuming a fundamental asymmetry in the interaction as a quark pair or diquark interacts with the third quark as a single quark.\cite{proposal}

However, the non-existence of the missing resonant states is not the only means of explaining the existing data. It is also possible that the missing states simply have not been observed. The existing experimental data are largely based on the analysis of reactions using pion beams. Resonances that couple less strongly to pions might have escaped detection, and it is possible that these resonances couple strongly to rarely explored channels. Indeed, several resonances are predicted to couple strongly to photons ($\gamma N$) and to kaons ($KY$).\cite{capstick}

The interaction of photon beams and nucleons is described by four complex amplitudes, which can only be fully determined when a complete set of measurements is performed, \emph{i.e.} when cross-section data are complemented by polarization data including beam, target and recoil asymmetries and combinations of beam-target, beam-recoil and target-recoil polarization asymmetries. This work specifically concerns the measurements of the $T$ and $F$ polarization observables, which are associated with a transversely polarized target and an unpolarized photon beam or circularly polarized photon beam, respectively. The cross section of $T$ and $F$ can be expressed as follows:

\small
\begin{equation}\label{FT-eqn}
\frac{d \sigma}{d \Omega} = \sigma_0[1 + P_T P_C F\cos(\beta - \phi) + P_T T\sin(\beta - \phi)],
\end{equation}
\normalsize

where $\sigma_0$ is the unpolarized cross section, $P_T$ is the degree of transverse polarization of the target and $\beta$ its orientation, $P_C$ is the circular polarization of the photon beam, and $\phi$ is the kaon azimuthal angle.

\section {Experimental Setup}

The FROST (FROzen Spin Target) experiment was performed at Jefferson Lab (JLab) in the spring and summer of 2010. The FROST target consists of butanol cooled to a temperature of approximately 30 mK by means of a $^{3}$He/$^{4}$He dilution refrigerator in order to maintain polarization with long relaxation times.\cite{frost} The decay products of the $K^{+}\Lambda$ and $K^{+}\Sigma^{0}$ final states were detected in the CEBAF Large Acceptance Spectrometer (CLAS) in Hall B of JLab\cite{clas-nim} and we can gather data on all single and double polarization observables.
The incident photon beam, from 0.5 to 2.4 GeV, was produced by means of bremsstrahlung from high-energy electrons; the interaction time and energy of the deflected electrons were measured in a high-resolution tagging system.\cite{tagger-nim} Circularly polarized photons with up to 80\% polarization were produced from longitudinally polarized electrons on an amorphous radiator.

\section {Results and Conclusion}

In order to perform this analysis, the raw data were calibrated and four-vectors were reconstructed from the CLAS detector information. The $K^{+}\Lambda$ and $K^{+}\Sigma^{0}$ events were identified using vertex cuts and cuts on $\beta$ and momentum of detected particles in conjuction with missing-mass techniques. A proton and a kaon were identified by means of a plot of $\beta$ versus momentum. Using the invariant mass from the initial photon and target proton, we find the missing mass of the detected proton and kaon and make a loose cut around the missing $\pi^{-}$ from 0.05 to 0.30 GeV. From events remaining after the cut, the $\Lambda$ and $\Sigma^{0}$ peaks are readily identified from the missing mass off the kaons.
To isolate the background, we use the extracted $\Lambda$ and $\Sigma^{0}$ quantities and fit both peaks with Gaussians. After that, the background is then fit with a cubic polynomial and finally, the mass range (here from 1.05 to 1.3 GeV) is fit for each cos $\theta$ bin. After isolating those events from background, we can again use the extracted $\Lambda$ and $\Sigma^{0}$ quantities to determine the various polarization observables using the Fourier moment method. 
 

To this end, the signs of the polarization terms ($P_T, P_C$) in eq.(\ref{FT-eqn}) are explicitly written out, {\em e.g.} after summing over both photon helicity states:
\small
\begin{equation}\label{T-cs}
\sigma^{(+)}=\sigma_0 \left[1+ |P_T^{(+)}| \sin(\beta-\phi) \right] \quad ; \quad 
\sigma^{(-)}=\sigma_0 \left[1- |P_T^{(-)}| \sin(\beta-\phi) \right] \ ,
\end{equation} 
\normalsize
such that
\small
\begin{equation}\label{T-eq}
T=\frac{\sigma^{(+)} - \sigma^{(-)}}{\sigma^{(+)} + \sigma^{(-)}}
=\frac{2\ \left( N^{(-)} Z^{(+)}_1 - N^{(+)} Z^{(-)}_1 \right) }
{N^{(-)} P_T^{(-)} \left( H^{(+)}_0-H^{(+)}_2 \right) +N^{(+)} P_T^{(+)} \left( H^{(-)}_0-H^{(-)}_2 \right) } \ ,
\end{equation}
\normalsize

where $Z^{(\pm)}_n$ are the $n$-th $\sin$ moments and $H^{(\pm)}_n$ are the $n$-th $\cos$ moments for positive and negative target-polarization orientation, respectively. A similar expression can be obtained for the $F$ asymmetry.\cite{mike}

\begin{figure}[h!tb]
\begin{minipage}{0.33\textwidth}
\centerline{\includegraphics[width=0.8\textwidth]{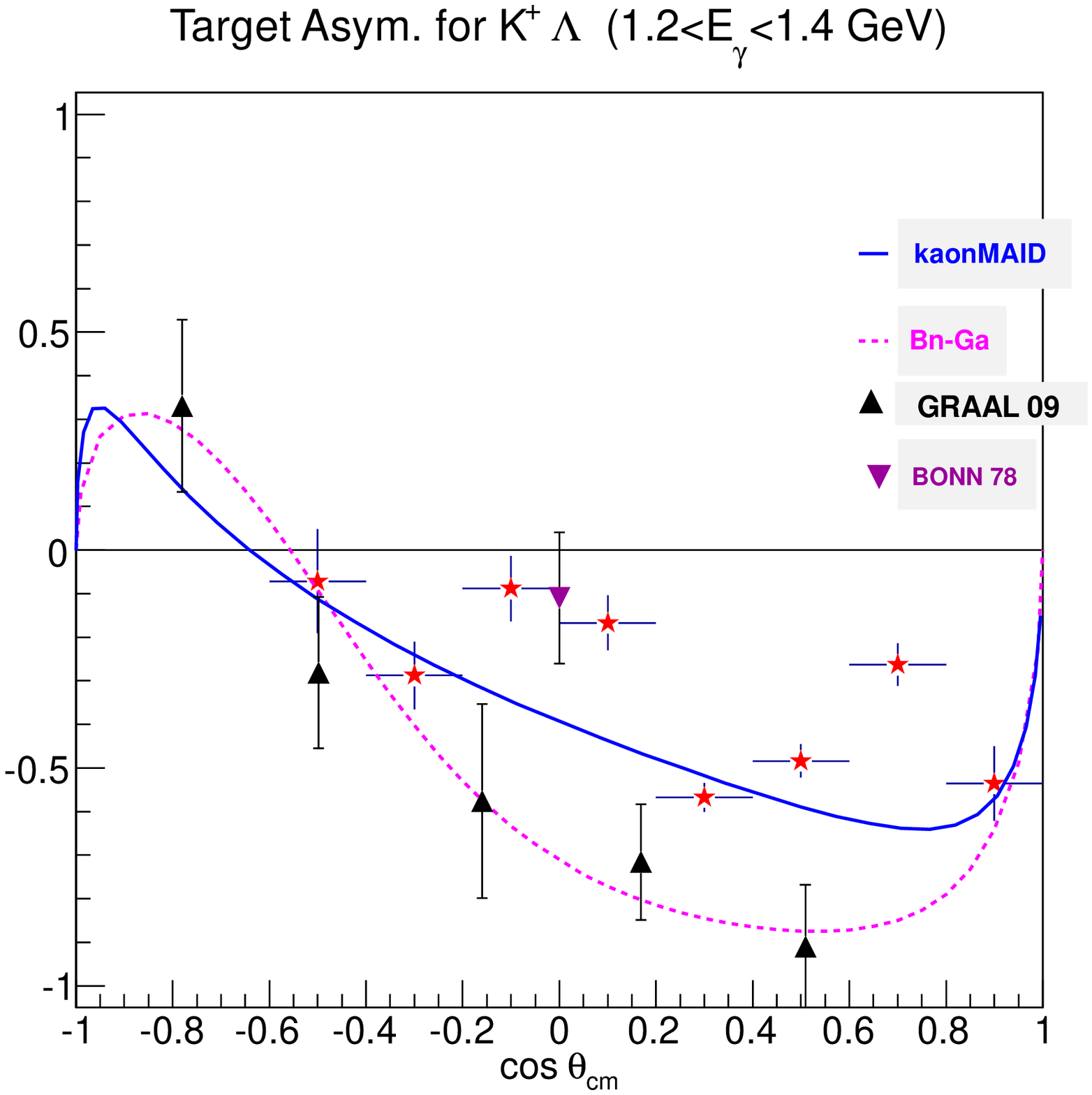}}
\end{minipage}\hfill
\begin{minipage}{0.33\textwidth}
\centerline{\includegraphics[width=0.8\textwidth]{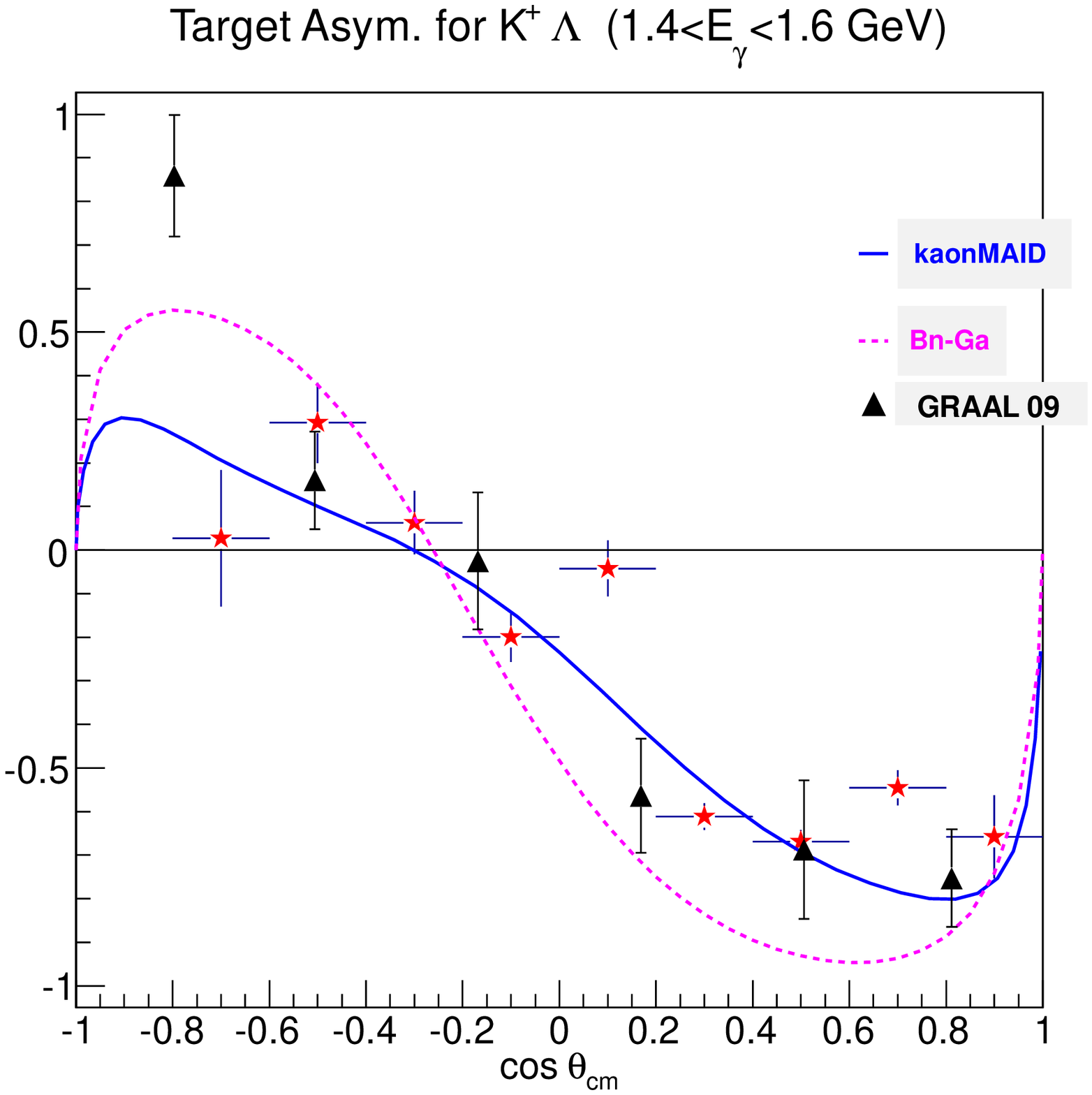}}
\end{minipage}\hfill
\begin{minipage}{0.33\textwidth}
\centerline{\includegraphics[width=0.8\textwidth]{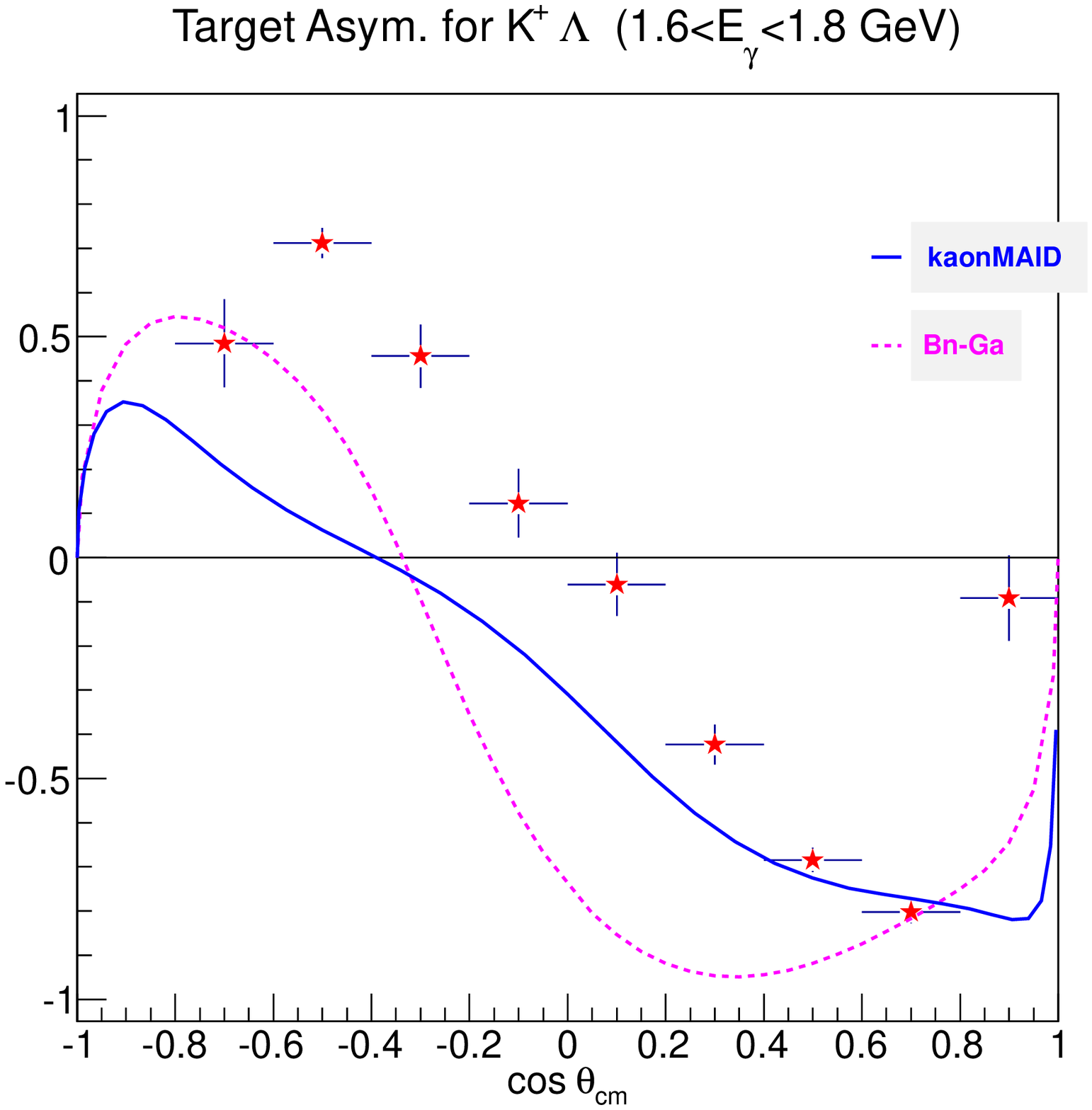}}
\end{minipage}\hfill
\caption{Preliminary CLAS results (red star) for $T$ asymmetry as a function of cos$\theta$ for $K^{+}\Lambda$. Overlaid on the plots are the current Bonn-Gatchina (dashed magenta) and kaonMAID (solid blue) PWA solutions (fit without this data). Also plotted with published results from GRAAL (2009) (black triangle) and Bonn (1978) (purple upside-down triangle). (Color online).
\label{f1}}
\end{figure}

Using all this information, we can finally construct the asymmetries for $T$ and $F$. Fig.~\ref{f1} shows preliminary results of $T$ for $K^{+}\Lambda$. The CLAS data (obtained through a direct measurement) are largely consistent with GRAAL and Bonn data, which were obtained through double-polarization data together with $O_{x'}$ and $O_{z'}$. 

\begin{figure}[h!tb]
\begin{minipage}{0.33\textwidth}
\centerline{\includegraphics[width=0.8\textwidth]{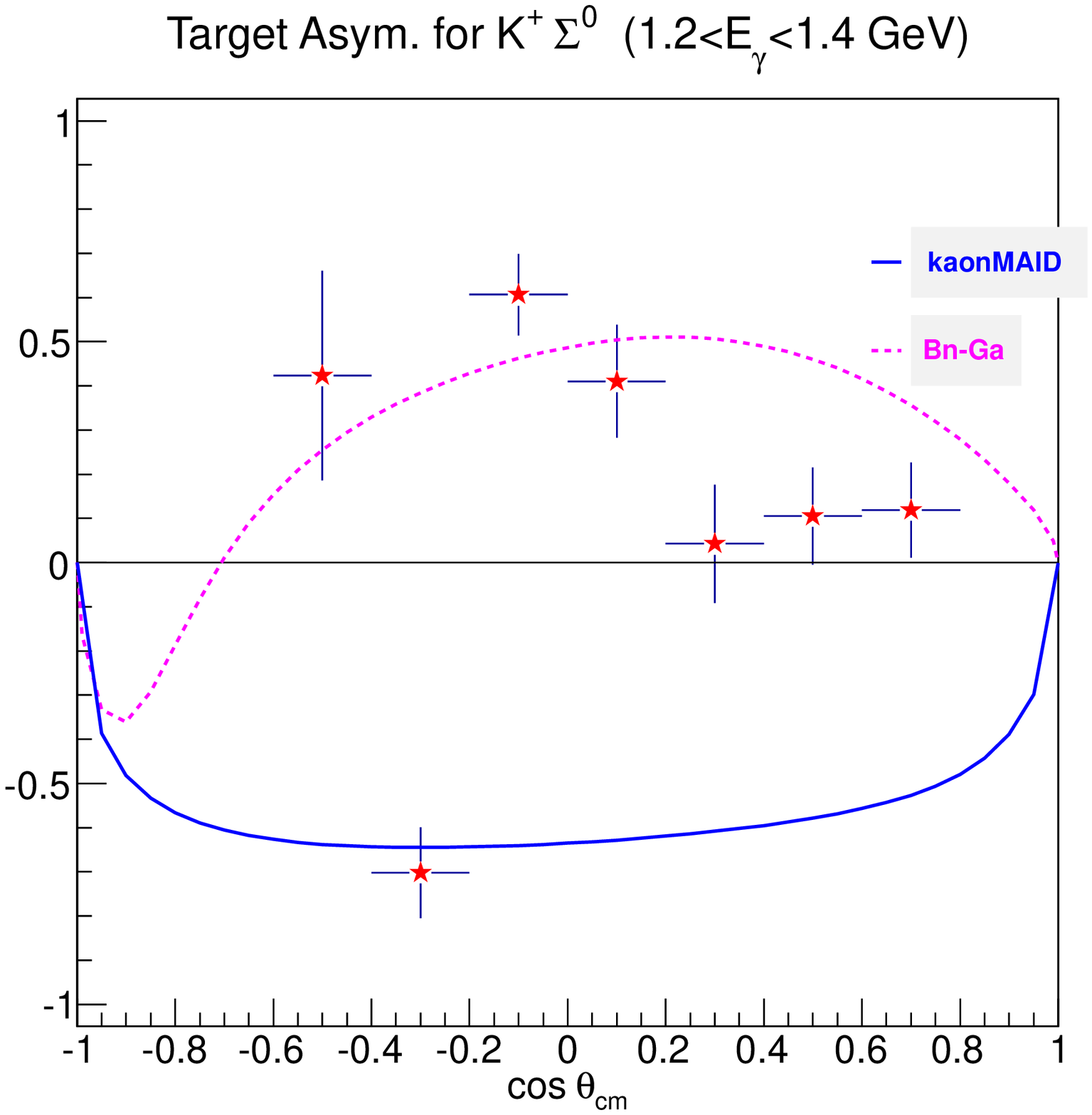}}
\end{minipage}\hfill
\begin{minipage}{0.33\textwidth}
\centerline{\includegraphics[width=0.8\textwidth]{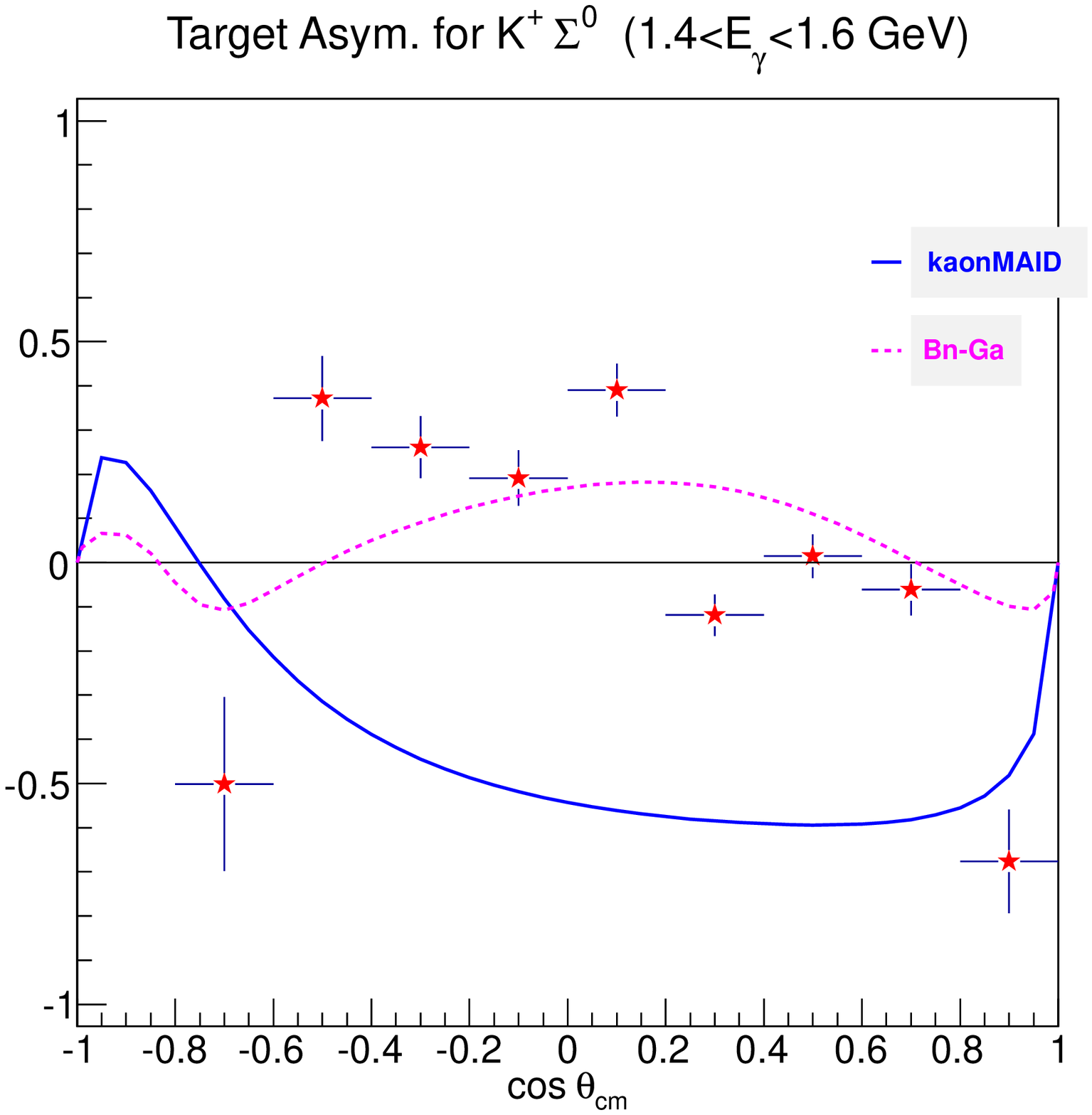}}
\end{minipage}\hfill
\begin{minipage}{0.33\textwidth}
\centerline{\includegraphics[width=0.8\textwidth]{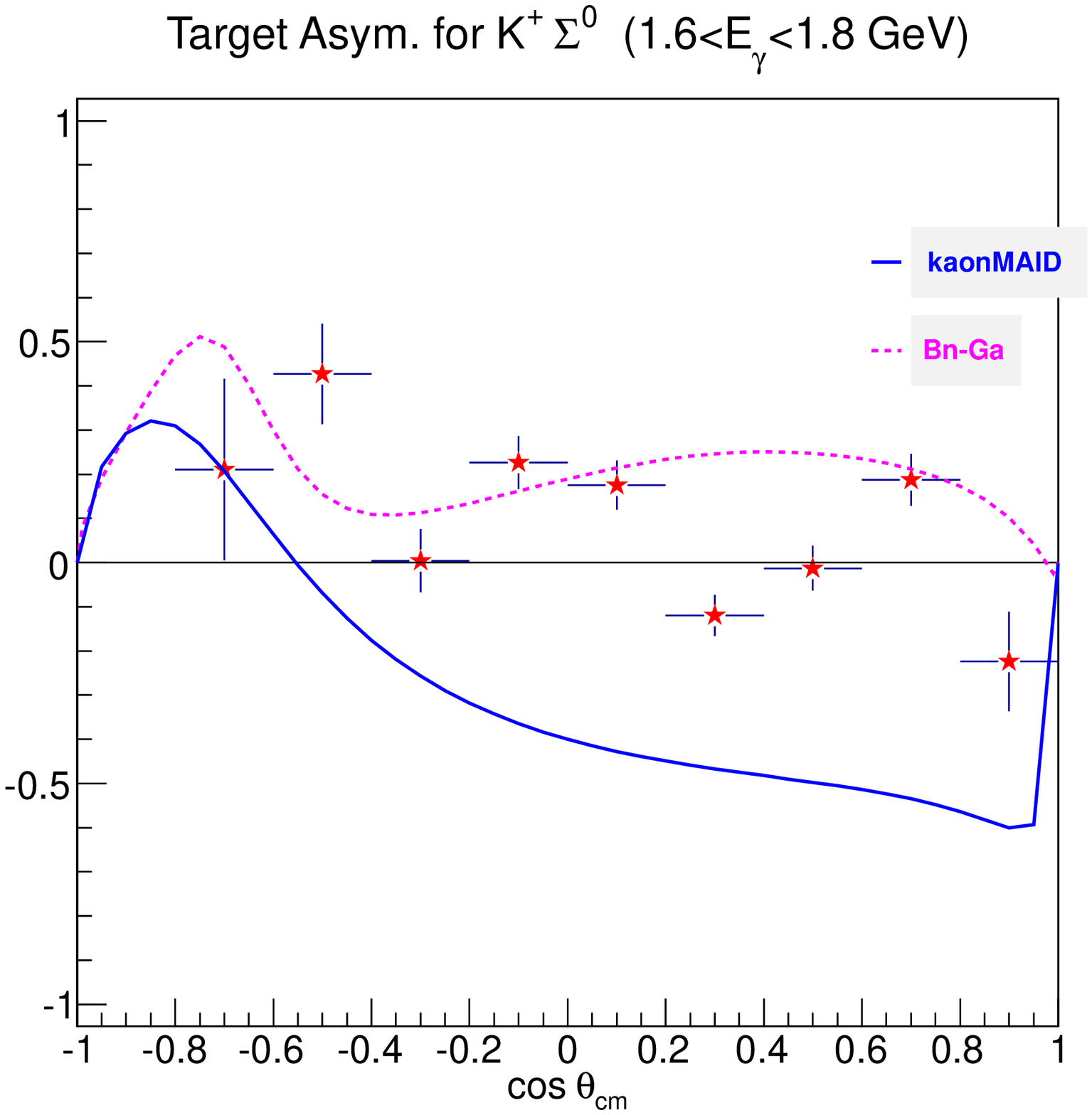}}
\end{minipage}\hfill
\caption{Preliminary CLAS results (red star) for $T$ asymmetry as a function of cos$\theta$ for $K^{+}\Sigma^0$. Overlaid on the plots are the current Bonn-Gatchina (dashed magenta) and kaonMAID (solid blue) PWA solutions (fit without this data). (Color online).
\label{f2}}
\end{figure}

Fig.~\ref{f2} shows preliminary results of $T$ for $K^{+}\Sigma^0$. There are no previously published data, but there is a descrepancy with PWA solutions at lower energies, but more consistent at higher energies, specifically the Bonn-Gatchina solutions. 

\begin{figure}[h!tb]
\begin{minipage}{0.33\textwidth}
\centerline{\includegraphics[width=0.8\textwidth]{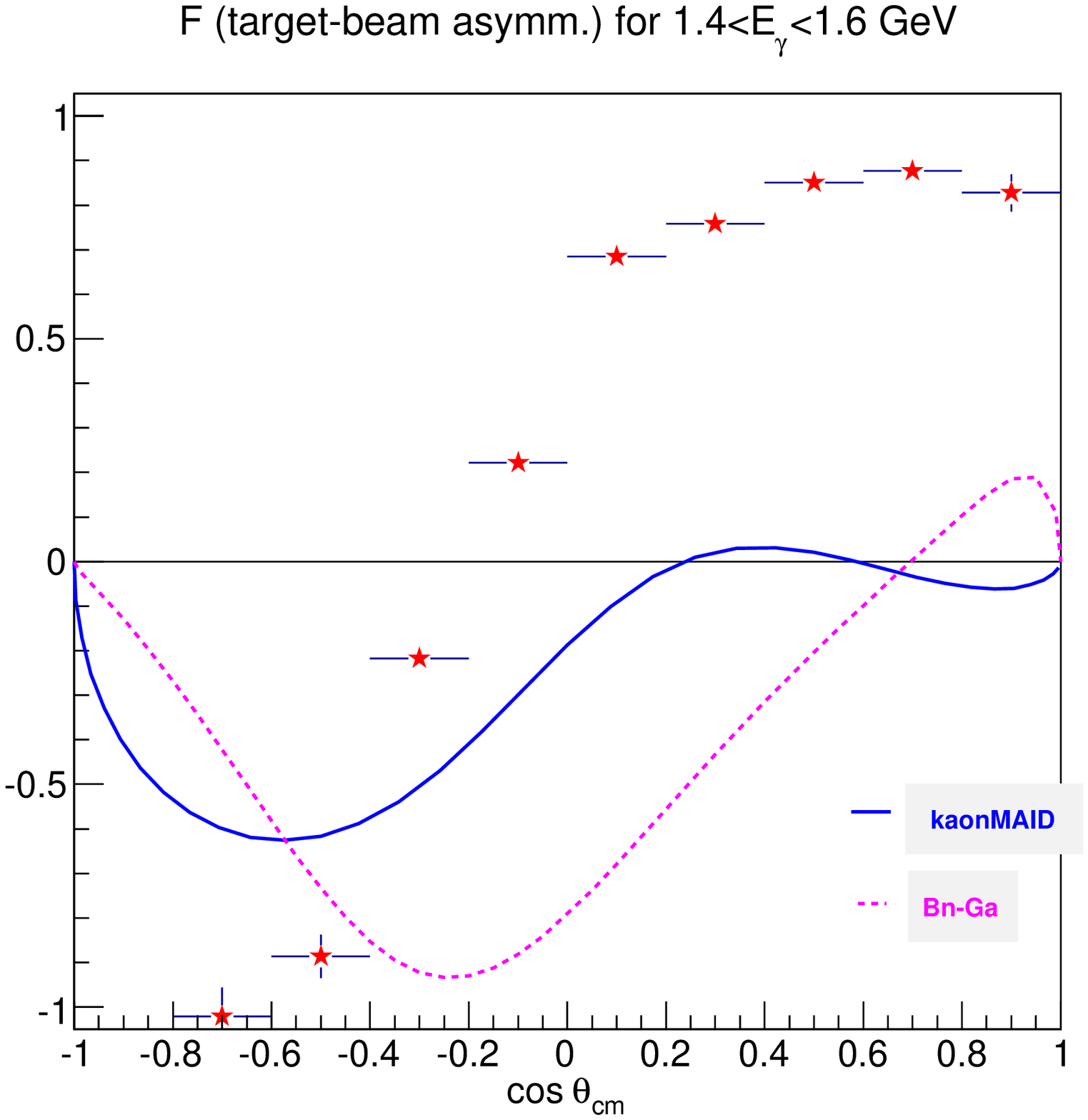}}
\end{minipage}\hfill
\begin{minipage}{0.33\textwidth}
\centerline{\includegraphics[width=0.8\textwidth]{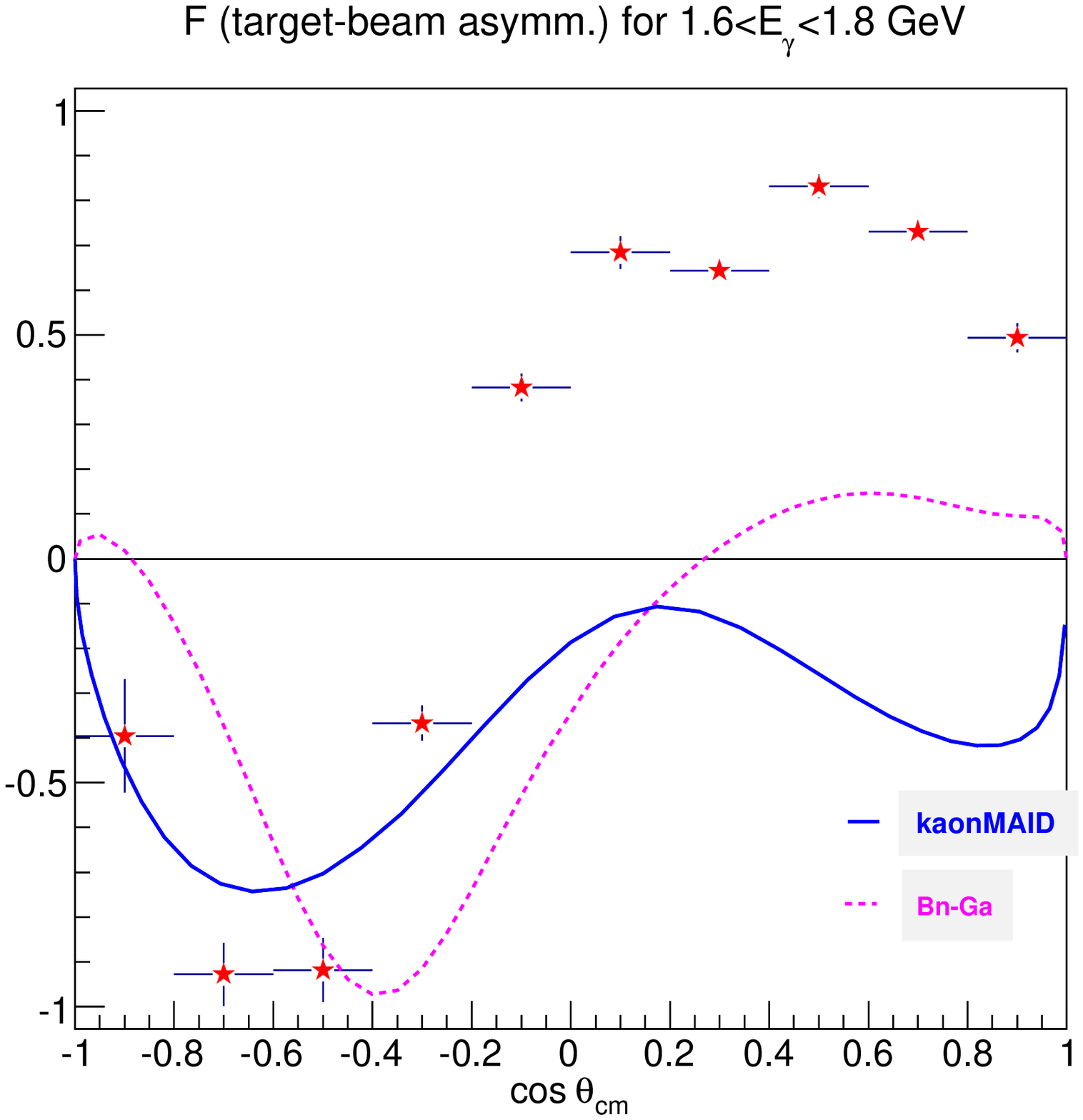}}
\end{minipage}\hfill
\begin{minipage}{0.33\textwidth}
\centerline{\includegraphics[width=0.8\textwidth]{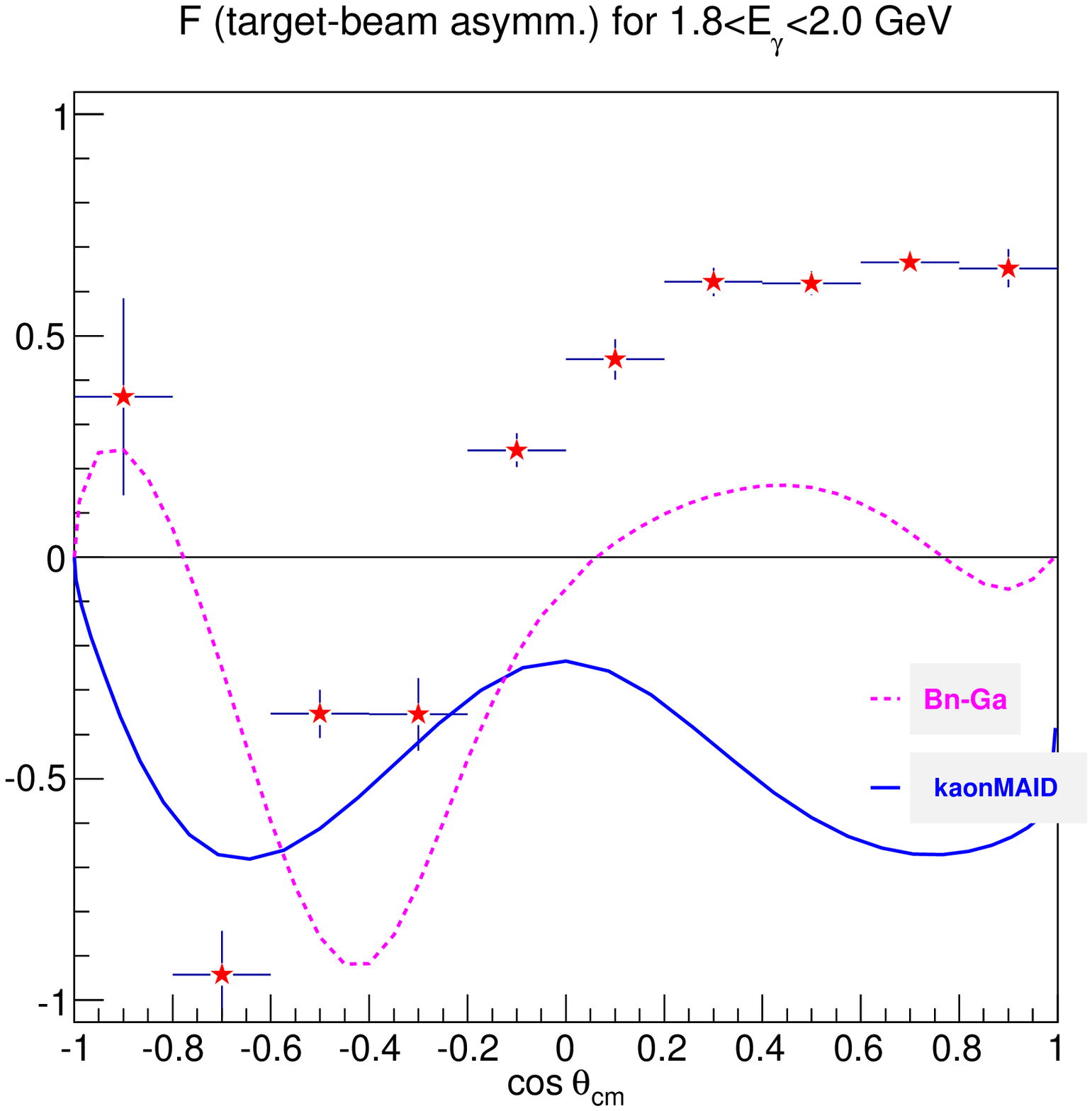}}
\end{minipage}\hfill
\caption{Preliminary CLAS results (red star) for $F$ asymmetry as a function of cos$\theta$ for $K^{+}\Lambda$. Overlaid on the plots are the current Bonn-Gatchina (dashed magenta) and kaonMAID (solid blue) PWA solutions (fit without this data). (Color online). 
\label{f3}}
\end{figure}

Also shown are preliminary results of $F$ for $K^{+}\Lambda$ in Fig.~\ref{f3}. There are no previously published data, but we can see that there are large descrepancies with the PWA solutions at all energies. 

In conclusion, it appears that this work can significantly increase the world database and change the current PWA solutions drastically. The CLAS results are ranging in energies never detected before and can provide long sought-out data that can help to constrain the various PWA solutions and then help to extract the undiscovered states of the nucleon. 

\section {Acknowledgments}

Our work at Jefferson Lab is supported in parts by the U.S. National Science Foundation: NSF PHY-0969434. Jefferson Science Associates operates the Thomas Jefferson National Accelerator Facility under DOE contract DE-AC05-84ER40150.


\end{document}